\documentclass[12pt]{article}
\usepackage{amsmath,amsfonts}
\makeatletter \@addtoreset{equation}{section}

\usepackage{cite}
\usepackage{bm}
\usepackage{dcolumn}
\makeatother
\def\one{{\hbox{ 1\kern-.8mm l}}}
\newcommand{\Dslash}{\not{\hbox{\kern-4pt $D$}}}
\newcommand{\pdslash}{\not{\hbox{\kern-2pt $\partial$}}}
\newcommand{\be}{\begin{equation}}
\newcommand{\bea}{\begin{eqnarray}}
\newcommand{\eea}{\end{eqnarray}}
\newcommand{\ba}{\begin{array}}
\newcommand{\ea}{\end{array}}
\newcommand{\ee}{\end{equation}}

\begin{document}

\begin{titlepage}
\vspace{10mm}

\vspace*{20mm}
\begin{center}
{\Large {\bf Criticality in Third Order Lovelock Gravity and Butterfly effect }\\
}
\vspace*{15mm}
\vspace*{1mm}
{Mohammad M. Qaemmaqami}

\vspace*{1cm}

{ \it  School of Particles and Accelerators\\Institute for Research in Fundamental Sciences (IPM)\\
	P.O. Box 19395-5531, Tehran, Iran }

\vspace*{0.5cm}
{E-mail: m.qaemmaqami@ipm.ir}%

\vspace*{2cm}

\end{center}

\begin{abstract}
We study third order Lovelock Gravity in $ D=7 $ at the critical point which three (A)dS vacua degenerate into one. We see there is not propagating graviton at the critical point. And also we compute the butterfly velocity for this theory at the critical point by considering the shock wave solutions near horizon, this is important to note that although there is no propagating graviton at the critical point, due to boundary gravitons the butterfly velocity is non-zero. Finally we observe that the butterfly velocity for third order Lovelock Gravity at the critical point in $ D=7 $ is less than the butterfly velocity for Einstein-Gauss-Bonnet Gravity at the critical point in $ D=7 $ which is less than the butterfly velocity in D = 7 for Einstein Gravity,\\ $  v_{B}^{E.H}>v_{B}^{E.G.B}>v_{B}^{3rd\,\,Lovelock} $. Maybe we can conclude that by adding higher order curvature corrections to Einstein Gravity the butterfly velocity decreases.

\end{abstract}

\end{titlepage}

\section{Introduction}
Introducing higher-derivative terms to Einstein Gravity has some advantages. Stelle had shown that extension of Einstein Gravity with higher order curvature invariants in four dimensions can be renormalizable \cite{Stelle:1976gc},\cite{Stelle:1977ry}, however, the theory contains ghost-like massive spin-2 modes. When there is a cosmological constant there exists a critical point in the parameter space of the coupling constants \cite{Lu:2011zk},\cite{Deser:2011xc} for which the ghost-like massive graviton becomes a logarithmic mode\cite{Bergshoeff:2011ri},\cite{Grumiller:2008qz},\cite{Alishahiha:2011yb},\cite{Johansson:2012fs},\cite{Alishahiha:2014dma},\cite{Alishahiha:2015whv}.\\
\\
In higher dimensions, there are some higher order curvature extensions for which equations of motion involve only two derivatives, and hence ghost excitations can be absent. These are Gauss-Bonnet or Lovelock Gravities \cite{Lovelock:1971yv}. In general, there are two (A)dS vacua in Einstein-Gauss-Bonnet Gravity. At the critical point, the two (A)dS vacua degenerate into one and the effective coupling for the kinetic term vanishes\cite{Fan:2016zfs}. The theory at the critical point does not have propagators therefore the linear modes are not gravitons and we have a theory of gravity without graviton.\\
\\
In general, there are three (A)dS vacua in third order Lovelock Gravity, at the critical point all (A)dS vacua degenerate into one. In section 3 we study linearized gravity and find that the effective coupling for the kinetic term vanishes at the critical point, therefore the theory does not have propagators at the critical point, in other words in this point there is a theory of gravity without graviton.\\
\\
In section 4 we study wave solutions and we find that equations of motion is automatically satisfied by plugging the wave ansatz at the critical point. Therefore we see it is equal to the absence of gravitons at the critical point since the ansatz of wave solution is in the light cone gauge and solutions are satisfied for both linear and non-linear level\cite{Alishahiha:2011yb}. In section 5 we study the butterfly effect for 7-dimensional third order Lovelock Gravity at the critical point. It has been shown \cite{Shenker:2013pqa},\cite{Shenker:2013yza},\cite{Roberts:2014isa},\cite{Leichenauer:2014nxa} that chaos in thermal CFT may be described by shock wave near the horizon of an AdS black hole. In other words the propagation of the shock wave on the horizon provides a description of the butterfly effect in the dual field theory.\\
\\
For computing the butterfly velocity in third order Lovelock Gravity at the critical point we consider the shock wave ansatz in Kruskal coordinate and finally we find a second order differential equation for perturbations near the horizon, thus we can read the butterfly velocity at the critical point. It is important to note that although at the critical point the theory does not have propagating gravitons, due to boundary
gravitons, the butterfly velocity is non-zero. Finally, we observe that the butterfly velocity of third order Lovelock Gravity at the critical point in D = 7 is less than the butterfly velocity of Einstein-Gauss-Bonnet Gravity at  the critical point
in D = 7 which is less than the butterfly velocity of Einstein Gravity in D = 7. Maybe we can conclude that by adding higher order curvature corrections to Einstein Gravity the butterfly velocity decreases.

\section{Third Order Lovelock Gravity}
In this section, we consider Third Order Lovelock Gravity\cite{Hendi:2008wq},\cite{Dehghani:2009zzb},\cite{deBoer:2011wk},\cite{Zou:2013ix},\cite{Xu:2014tja},\cite{Konoplya:2017lhs},\cite{Banados:2005rz} without matter fields. The Lagrangian density is given by
\begin{eqnarray}
\mathcal{L}=\frac{1}{16\pi G_{N}} \sqrt{-g}\big[R-2\Lambda_{0}+\alpha_{2}\mathcal{L}_{2}+\alpha_{3}\mathcal{L}_{3}\big],
\end{eqnarray}
where $ \Lambda_{0} $ is the bare cosmological constant. $ \alpha_{2} $ and $ \alpha_{3} $ are coefficients of Gauss-Bonnet term $ \mathcal{L}_{2} $ and third order Lovelock term $ \mathcal{L}_{3} $ respectively and $ G_{N} $ is Newton's constant. The expression of $ \mathcal{L}_{2} $ and $ \mathcal{L}_{3} $ are
\begin{eqnarray}
&&\mathcal{L}_{2}= R^{2}-4R_{\mu\nu}R^{\mu\nu}+R_{\mu\nu\rho\sigma}R^{\mu\nu\rho\sigma},\\\nonumber
&&\mathcal{L}_{3}= 2R^{\mu\nu\sigma\kappa}R_{\sigma\kappa\rho\tau}R^{\rho\tau}~_{\mu\nu}+8R^{\mu\nu}~_{\sigma\rho}R^{\sigma\kappa}~_{\nu\tau}R^{\rho\tau}~_{\mu\kappa}+24R^{\mu\nu\sigma\kappa}R_{\sigma\kappa\nu\rho}R^{\rho}~_{\mu}+\\\nonumber &&+3RR^{\mu\nu\sigma\kappa}R_{\mu\nu\sigma\kappa}+24R^{\mu\nu\sigma\kappa}R_{\sigma\mu}R_{\kappa\nu}+16R^{\mu\nu}R_{\nu\sigma}R^{\sigma}~_{\mu}-12RR^{\mu\nu}R_{\mu\nu}+R^{3}.
\end{eqnarray}
Varying the action with respect to metric, the equation of motion is:
\begin{eqnarray}
E_{\mu\nu}=R_{\mu\nu}-\frac{1}{2}Rg_{\mu\nu}+\Lambda_{0} g_{\mu\nu}+\alpha_{2}G^{(2)}_{\mu\nu}+\alpha_{3}G^{(3)}_{\mu\nu}=0,
\end{eqnarray}
where $ G^{(2)}_{\mu\nu} $ and $ G^{(3)}_{\mu\nu} $ are the Gauss-Bonnet and third order Lovelock tensors respectively:
\begin{eqnarray}
G^{(2)}_{\mu\nu}&=&2\big(R_{\mu\sigma\kappa\tau}R_{\nu}~^{\sigma\kappa\tau}-2R_{\mu\rho\nu\sigma}R^{\rho\sigma}-2R_{\mu\sigma}R^{\sigma}~_{\nu}+RR_{\mu\nu}\big)- \frac{1}{2}\mathcal{L}_{2}g_{\mu\nu},\\\nonumber
G^{(3)}_{\mu\nu}&=&3R_{\mu\nu}R-12RR_{\mu\sigma}R^{\sigma}~_{\nu}-12R_{\mu\nu}R_{\alpha\beta}R^{\alpha\beta}+24R_{\mu}~^{\alpha}R_{\alpha}~^{\beta}R_{\beta\nu}-24R_{\mu}~^{\alpha}R^{\beta\sigma}R_{\alpha\beta\sigma\nu}+\\\nonumber
&&+3R_{\mu\nu}R_{\alpha\beta\sigma\kappa}R^{\alpha\beta\sigma\kappa}-12R_{\mu\alpha}R_{\nu\beta\sigma\kappa}R^{\alpha\beta\sigma\kappa}-12RR_{\mu\sigma\nu\kappa}R^{\sigma\kappa}+6RR_{\mu\alpha\beta\sigma}R_{\nu}~^{\alpha\beta\sigma}+\\\nonumber
&&+24R_{\mu\alpha\nu\beta}R_{\sigma}~^{\alpha}R^{\sigma\beta}+24R_{\mu\alpha\beta\sigma}R_{\nu}~^{\beta}R^{\alpha\sigma}+24R_{\mu\alpha\nu\beta}R_{\sigma\kappa}R^{\alpha\sigma\beta\kappa}-12R_{\mu\alpha\beta\sigma}R^{\kappa\alpha\beta\sigma}R_{\kappa\nu}- \\\nonumber
&&-12R_{\mu\alpha\beta\sigma}R^{\alpha\kappa}R_{\nu\kappa}~^{\beta\sigma}+24R_{\mu}~^{\alpha\beta\sigma}R_{\beta}~^{\kappa}R_{\sigma\kappa\nu\alpha}-12R_{\mu\alpha\nu\beta}R^{\alpha}~_{\sigma\kappa\rho}R^{\beta\sigma\kappa\rho}-\\\nonumber
&&-6R_{\mu}~^{\alpha\beta\sigma}R_{\beta\sigma}~^{\kappa\rho}R_{\kappa\rho\alpha\nu}-24R_{\mu\alpha}~^{\beta\sigma}R_{\beta\rho\nu\lambda}R_{\sigma}~^{\lambda\alpha\rho}- \frac{1}{2}\mathcal{L}_{3}g_{\mu\nu}.
\end{eqnarray}

If we put (A)dS space-time values for curvatures(maximally symmetric relations for curvatures) in terms of metric in D-dimensions:
\begin{eqnarray}
R_{\mu\nu\lambda\sigma}=\frac{2\Lambda}{(D-1)(D-2)}\big(g_{\mu\lambda}g_{\nu\sigma}-g_{\mu\sigma}g_{\nu\lambda}\big),  \,\,\,\,\      \,\,\,\,\  R_{\mu\nu}=\frac{2\Lambda}{D-2}g_{\mu\nu},  \,\,\,\,\     \,\,\,\,\   R=\frac{2D\Lambda}{D-2},
\end{eqnarray}
 the effective cosmological constant $ \Lambda $ satisfies a cubic algebraic equation:
\begin{eqnarray}
\frac{1}{2}(\Lambda-\Lambda_{0})+\frac{(D-3)(D-4)}{(D-1)(D-2)}\alpha_{2}\Lambda^{2}+2\frac{(D-3)(D-4)(D-5)(D-6)}{(D-1)^{2}(D-2)^{2}}\alpha_{3}\Lambda^{3}=0.
\end{eqnarray}
Here we consider specially $ D=7 $, The above equation for effective cosmological constant $ \Lambda $ in $ D=7 $ becomes:
\begin{eqnarray}
75(\Lambda-\Lambda_{0})+60\alpha_{2}\Lambda^{2}+8\alpha_{3}\Lambda^{3}=0.
\end{eqnarray}
One can see for $ D=7 $, all three solutions of the above equation degenerate into one for $ \alpha_{3}=2\alpha_{2}^{2} $, we call this point the critical point\cite{Crisostomo:2000bb}, in this point all solutions degenerate into $ \Lambda=-\frac{5}{4\alpha_{2}} $ and $ \Lambda_{0}=-\frac{5}{12\alpha_{2}} $ or $ \Lambda=3\Lambda_{0} $, H. Lu et.al studied the critical point for Einstein-Gauss-Bonnet Gravity\cite{Fan:2016zfs} and they find $ \Lambda =2\Lambda_{0} $ for critical point. If we assume AdS space-time for solution in $ D=7 $, when they degenerate into one  at the critical point we have:
\begin{eqnarray}
 \Lambda =3\Lambda_{0}=-\frac{15}{l^{2}}, \,\,\,\,\   \,\,\,\,\,\,\,\,\,\  \alpha_{2}=\frac{l^{2}}{12},    \,\,\,\,\  \,\,\,\,\,\,\,\,\,\    \alpha_{3}=2\alpha_{2}^{2}=\frac{l^{4}}{72},
\end{eqnarray}
  where $ l $ is the AdS radius.
\section{Linearized gravity}
In this section we like to study the spectrum of perturbation around (A)dS vacua in third order Lovelock Gravity in $ D=7 $, at the critical point eq(2.8). To do this, let us parametrize the perturbation as below:
\begin{eqnarray}
g_{\mu\nu}=\bar{g}_{\mu\nu}+h_{\mu\nu},
\end{eqnarray}
where $ \bar{g}_{\mu\nu} $ means the (A)dS vacua for general parameters. By using the linearized Einstein tensor around (A)dS vacuum in $ D=7 $
\begin{eqnarray}
\mathcal{G}^{L}_{\mu\nu}=R^{L}_{\mu\nu}- \frac{1}{2}\bar{g}_{\mu\nu}R^{L}- \frac{2\Lambda}{5}h_{\mu\nu},
\end{eqnarray} 
where the linearized Ricci tensor $ R^{L}_{\mu\nu} $ and Ricci scalar $ R^{L} $ in $ D=7 $ are respectively
\begin{eqnarray}
R^{L}_{\mu\nu}=\frac{1}{2}\bigg(\bar{\nabla}^{\sigma}\bar{\nabla}_{\mu}h_{\nu\sigma}+\bar{\nabla}^{\sigma}\bar{\nabla}_{\nu}h_{\mu\sigma}-\bar{\Box}h_{\mu\nu}-\bar{\nabla}_{\mu}\bar{\nabla}_{\nu}h\bigg), \,\,\    \,\,\   R^{L}=-\bar{\Box}h+\bar{\nabla}^{\sigma}\bar{\nabla}^{\mu}h_{\mu\sigma}- \frac{2\Lambda}{5}h
\end{eqnarray}
We obtain the linearized equations of motion for this model
\begin{eqnarray}
\kappa_{eff}\mathcal{G}^{L}_{\mu\nu}=0, \,\,\,\,\,\,\,\              \,\,\,\,\,\,\,\  \kappa_{eff}=25+40\alpha_{2}\Lambda+8\alpha_{3}\Lambda^{2}.
\end{eqnarray}
With the effective cosmological constant being $ 3\Lambda_{0} $ at the critical point in above equation (3.2) we have $ \kappa_{eff}=0 $ and hence the linearized equations of motion are automatically satisfied.\\
\\
At the critical point the linearized equations of motion in the above section are automatically satisfied. And also there is no kinetic term for the fluctuation $ h_{\mu\nu} $ at the quadratic order, therefore the theory does not have any propagator, and hence it is no longer proper to take $ h_{\mu\nu} $ as usual graviton modes. We have thus a theory of gravity without graviton! We arrived at the above critical point by studying the linearized equations of the third order Lovelock Gravity. It happens that at the critical point the theory also admits only one AdS vacuum.
\section{Critical point and wave Solutions}
In last section we observed that third order Lovelock Gravity at the critical point eq(2.8) does'nt contain any propagating mode at the linearized level. The interesting question which one can ask is that this model at the critical point admit propagating modes at non-perturbative level.\\
\\
To answer this question, one can study the wave solutions, we can write the ansatz for AdS wave solutions as follows\cite{Alishahiha:2011yb}:
\begin{eqnarray}
g_{\mu\nu}=\bar{g}_{\mu\nu}+F(r,u)k_{\mu}k_{\nu},
\end{eqnarray}
where $ k_{\mu} $ is a null vector field($ k_{\mu}k^{\mu}=0 $) with respect to the background metric $ \bar{g}_{\mu\nu} $, which is $ AdS_{7} $ metric, then the solutions are satisfied for both linear and non-linear level.\\
\\
Note that F is independent of the integral parameter along $ k_{\mu} $ therefore one can write the ansatz as follows\cite{Alishahiha:2011yb}:
\begin{eqnarray}
ds^{2}=\frac{L^{2}}{r^{2}}\big[-F(u,r)du^{2}-2dudv+dr^{2}+dx_{i}dx^{i}\big], \,\,\,\,\       \,\,\,\,\  i=x,y,z,w.
\end{eqnarray}
If we plug the above ansantz in the equations of motion(Eq(2.3)) from $ uv, rr, xx, yy, zz, ww $ components we find:
\begin{eqnarray}
75(\Lambda-\Lambda_{0})+60\alpha_{2}\Lambda^{2}+8\alpha_{3}\Lambda^{3}=0.
\end{eqnarray}
That is exactly the equation for effective cosmological constant $ \Lambda $ in $ D=7 $ eq(2.7), The other non-vanishing component of equations of motion which is $ uu $ component is:
\begin{eqnarray}
&&-10\big(75(\Lambda-\Lambda_{0})+60\alpha_{2}\Lambda^{2}+8\alpha_{3}\Lambda^{3}\big)F(u,r)+\\\nonumber
&&+\Lambda\big(25+40\alpha_{2}\Lambda+8\alpha_{3}\Lambda^{2}\big)\bigg(r^{2}\frac{\partial^{2} F(u,r)}{\partial r^{2}}-5r\frac{\partial F(u,r)}{\partial r}\bigg)=0.
\end{eqnarray}
The last equation by using eq(4.3) simplified as:
\begin{eqnarray}
\Lambda\big(25+40\alpha_{2}\Lambda+8\alpha_{3}\Lambda^{2}\big)\bigg(r^{2}\frac{\partial^{2} F(u,r)}{\partial r^{2}}-5r\frac{\partial F(u,r)}{\partial r}\bigg)=0.
\end{eqnarray}
For $ \Lambda\neq 0 $ and $ 25+40\alpha_{2}\Lambda+8\alpha_{3}\Lambda^{2}\neq 0 $ the solution is:
\begin{eqnarray}
F(u,r)=\frac{c_{1}(u)}{6}r^{6}+c_{2}(u).
\end{eqnarray}
But we know from eq(3.4), at the critical point therefore the overall factor of the differential equation eq(4.5) is zero therefore equations of motion are automatically satisfied by plugging the wave ansatz. Note that for the above ansatz of wave solution which is in light cone gauge the solutions are satisfied for both linear and non-linear level. Therefore the above result about equations of motion at the critical point which are automatically satisfied by plugging the wave ansatz, is equal to absence of graviton at the critical point in third order Lovelock Gravity.
\section{Shock wave in third order Lovelock Gravity at the critical point}
In this section, we study butterfly effect in 7-dimensional third order Loveock Gravity at the critical point. Indeed it was shown \cite{Shenker:2013pqa},\cite{Shenker:2013yza},\cite{Roberts:2014isa},\cite{Leichenauer:2014nxa} the propagation of the shock wave on the horizon of an AdS black hole provides a description of butterfly effect in the dual field theory.\\
\\
In field theory side butterfly effect may be diagnosed by out of time order four point function between pairs of local operators
\begin{eqnarray}
\langle V_{x}(0)W_{y}(t)V_{x}(0)W_{y}(t)\rangle_{\beta},
\end{eqnarray}
where $ \beta $ is inverse of the temperature. The butterfly effect may be seen by a sudden decay after the scrambling time, $ t_{*} $ ,
\begin{eqnarray}
\frac{\langle V_{x}(0)W_{y}(t)V_{x}(0)W_{y}(t)\rangle_{\beta}}{\langle V_{x}(0)V_{x}(0)\rangle_{\beta} \langle W_{y}(t)W_{y}(t)\rangle_{\beta}}\sim 1-e^{\lambda_{L}\big(t-t_{*}-\frac{|x-y|}{v_{B}}\big)},
\end{eqnarray}
where $ \lambda_{L} $ is the Lyapunov exponent and $ v_{B} $ is the butterfly velocity. The Lyapunov exponent is, $ \lambda_{L}=\frac{2\pi}{\beta} $, where $ \beta $ is inverse of Hawking temperature. And also the butterfly velocity should be identified by the velocity of shock wave by which the perturbation spreads in the space.\\
\\
To study the butterfly effect, we consider the black brane solution. The equations of motion of third order Lovelock Gravity at the critical point admit this asymptotically AdS black brane solution 
\begin{eqnarray}
ds^{2}=-f(r)dt^{2}+\frac{dr^{2}}{f(r)}+\frac{r^{2}}{l^{2}}d\vec{x}^{2},    \,\,\,\,\    \,\,\,\,\    f(r)=\frac{r^{2}}{l^{2}}\bigg(1-\frac{r_{h}^{2}}{r^{2}}\bigg),
\end{eqnarray}
where $ r_{h} $ is the radius of horizon.\\
\\
Now the aim is to study shock wave of this model when the above black hole solution of this theory is perturbed by injection of a small amount of energy. For this aim, it is better to rewrite the solution in Kruskal coordinate
\begin{eqnarray}
u=exp\big[\frac{2\pi}{\beta}(r_{*}-t)\big],  \,\,\,\,\    \,\,\,\,\   v=-exp\big[\frac{2\pi}{\beta}(r_{*}-t)\big],
\end{eqnarray}
where $ \beta=\frac{4\pi}{f^{'}(r)} $ is the inverse of temperature and $ dr_{*}=\frac{dr}{f(r)} $ is the tortoise coordinate.\\
\\
By making use of this coordinate system, the metric becomes into this form\cite{Shenker:2013pqa},\cite{Alishahiha:2016cjk}:
\begin{eqnarray}
ds^{2}=2A(uv)dudv+B(uv)d\vec{x}^{2}.
\end{eqnarray}
Here $ A(uv) $ and $ B(uv) $ are two functions, given by $ f(r) $, whose near horizon expansions are
\begin{eqnarray}
A(x)&=&-2cl^{2}\bigg(1-2cx+3c^{2}x^{2}-4c^{3}x^{3}+...\bigg),\\\nonumber
B(x)&=&\frac{r_{h}^{2}}{l^{2}}\bigg(1-4cx+8c^{2}x^{2}-12c^{3}x^{3}+...\bigg),
\end{eqnarray}
where $ c $ is an integration constant. Now we must study the shock wave, for this aim let us consider an injection of a small mount of energy from boundary toward the horizon at time $ -t_{w} $. This will cross the $ t=0 $ time slice while it is red shifted. Therefore the equations of motion should be deformed as 
\begin{eqnarray}
\mathcal{E}_{\mu\nu}=\kappa T^{s}_{\mu\nu},
\end{eqnarray}
where $ \kappa=8\pi G_{N} $, the energy-momentum tensor has only $ uu $ component due to energy injection:
\begin{eqnarray}
T_{uu}^{S}=lE\bigg(exp\big(\frac{2\pi t_{w}}{\beta}\big)\delta(u)\delta^{5}(\vec{x})\bigg).
\end{eqnarray}
For solving the equations of motion near horizon to find the shock wave solution, we consider this ansatz for back-reacted geometry
\begin{eqnarray}
ds^{2}=2A(UV)dUdV+B(UV)d\vec{x}^{2}-2A(UV)h(\vec{x})\delta(U)dU^{2},
\end{eqnarray} 
where the new coordinate $ U $ and $ V $ are
\begin{eqnarray}
U\equiv u,   \,\,\,\,\     \,\,\,\,\  V\equiv v+h(\vec{x})\Theta(u).
\end{eqnarray}
Plugging the ansatz into the equations of motion, near horizon at the leading order one finds a second order differential equation for $ h(\vec{x}) $
\begin{eqnarray}
\bigg(\frac{l^{2}}{r_{h}^{2}}\partial_{i}\partial^{i}-\frac{5}{l^{2}}\bigg)h(x^{i})=-\frac{1}{2cl^{2}}\big[\kappa lEe^{2\pi t_{w}/\beta}\big]\delta^{5}(x^{i}),
\end{eqnarray}
we can reduce the equation of motion into:
\begin{eqnarray}
(\partial_{i}\partial^{i}-a^{2})h(x^{i})=b\delta^{5}(x^{i}),   \,\,\,\,\,\,\,\,\,\      a^{2}=\frac{5r_{h}^{2}}{l^{4}},   \,\,\,\,\,\,\,\,\,\      b=-\frac{r_{h}^{2}}{2cl^{4}}\big[\kappa lEe^{2\pi t_{w}/\beta}\big],
\end{eqnarray}
By making use from planar symmetry of $ x^{i} $ coordinate directions we can choose one direction for solving the above differential. For this aim we assume injecting energy along one direction $ x $, then the energy-momentum is $ T_{uu}^{S}=lE\big(exp(\frac{2\pi t_{w}}{\beta})\delta(u)\delta(x)\big) $ therefore the differential equation reduces to
\begin{eqnarray}
\big(\partial_{i}\partial^{i}-a^{2}\big)h(x)=b\delta(x),
\end{eqnarray}
 whose solution is
\begin{eqnarray}
h(x)=-\frac{b}{2a}e^{-a|x|}
\end{eqnarray}
By replacing the values of $ a $ and $ b $, one can see $ h(x)\propto e^{\frac{2\pi}{\beta}\big[(t_{w}-t_{*})-|x|/v_{B}\big]} $, where the scrambling time is $ t_{*}=\frac{\beta}{2\pi}log(\frac{l^{5}}{\kappa}) $, with $ \kappa=8\pi G_{N} $ and $ G_{N} $ is Newton's constant in $ D=7 $, therefore from the above relation eq(5.14) we have $ a=\frac{2\pi}{\beta}\frac{1}{v_{B}} $\cite{Roberts:2014isa},\cite{Alishahiha:2016cjk}, then one can read the value of butterfly velocity at the critical point:
\begin{eqnarray}
v_{B}=\frac{2\pi}{\beta a}=\sqrt{\frac{1}{5}},   \,\,\,\,\    \,\,\,\,\    \frac{2\pi}{\beta}=\frac{f^{'}(r)}{2}=\frac{r_{h}}{l^{2}}.
\end{eqnarray}
It has worth to note that in this case although at the critical point the theory has not propagating gravitons, due to boundary gravitons the butterfly velocity is non-zero. And also if we compare this velocity with butterfly velocity for Eeistein-Gauss-Bonnet Gravity at critical point in $ D=7 $ \cite{Roberts:2014isa},\cite{Alishahiha:2016cjk} ,$ v_{B}^{E.G.B}=\sqrt{\frac{3}{10}} $\big(in D-dimension at critical point $ v_{B}^{E.G.B}=\sqrt{\frac{D-1}{4(D-2)}} $\big), we observe butterfly velocity for Einstein-Gauss-Bonnet Gravity at critical point is larger than butterfly velocity for third order Lovelock Gravity at critical point. In addition butterfly velocity for Einstein Gravity\cite{Shenker:2013pqa}, in $ D=7 $ is $ v_{B}^{E.H}=\sqrt{\frac{3}{5}} $\big(in D-dimension $ v_{B}^{E.H}=\sqrt{\frac{D-1}{2(D-2)}} $\big), which is larger than butterfly velocity for Einstein-Gauss-Bonnet Gravity at critical point in $ D=7 $,
\begin{eqnarray}
 v_{B}^{E.H}>v_{B}^{E.G.B}>v_{B}^{3rd\,\,Lovelock}.
\end{eqnarray}
Maybe we can conclude that by adding higher order curvature corrections to Einstein Gravity the butterfly velocity decreases. In other word if we add more higher order curvature corrections to Einstein Gravity the butterfly velocity is less than butterfly velocity of lower order curvature one. 
\section{Conclusions}
In this paper we studied third order Lovelock Gravity at the critical point which all three (A)dS vacua degenerate into one. We observed that there is no propagating graviton at the critical point, in other words, in this point there is a theory of gravity without graviton. And also we studied wave solutions for this model and observed that the equations of motion are automatically satisfied and we know that for the wave solution ansatz in light cone gauge the solutions are satisfied for both linear and non-linear level, therefore it is equal to the absence of graviton at the critical point.\\
\\
In following we studied the butterfly effect for third order Lovelock Gravity at the critical point by considering the shock wave near horizon and we computed the butterfly velocity for this theory at the critical point in $ D=7 $. It is important to note that although at the critical point the theory has no propagating gravitons, due to boundary gravitons the butterfly velocity is non-zero. This is similar to what happens in three dimensional Einstein Gravity.\\
Finally we observe that the butterfly velocity of third order Lovelock Gravity at the critical point in $ D=7 $ is less than the butterfly velocity of Einstein-Gauss-Bonnet Gravity at the critical point in $ D=7 $ which is less than the the Butterfly velocity of Einstein Gravity in $ D=7 $. Maybe we can conclude that with adding higher order curvature corrections to Einstein Gravity the butterfly velocity decreases. Maybe decreasing the butterfly velocity by adding higher order curvatures to Einstein Gravity is related to changing the effective cosmological constant at the critical point by adding higher order curvatures, we know that the effective cosmological constant for third order Lovelock Gravity in $ D=7 $ at the critical point is $ \Lambda=3\Lambda_{0} $ and the effective cosmological constant for Einstein-Gauss-Bonnet Gravity in $ D=7 $ at critical point is $ \Lambda=2\Lambda_{0} $ \cite{Fan:2016zfs} and for Einstein Gravity we have $ \Lambda=\Lambda_{0} $. We see by adding higher order curvatures, the effective cosmological constant, $ \Lambda $,  increase in terms of the bare cosmological constant, $ \Lambda_{0} $. Maybe we can conclude that with increasing the effective cosmological constant in terms of bare cosmological constant, the butterfly velocity decreases. In other words maybe increasing the effective cosmological constant in terms of the bare cosmological constant means increasing an effective mass which causes the butterfly velocity to decrease. Recently people have done some investigations about the butterfly effect \cite{Feng:2017wvc},\cite{Baggioli:2017ojd},\cite{Cai:2017ihd},\cite{Ling:2016wuy},\cite{Kim:2017dgz},\cite{Qi:2017ttv},\cite{Blake:2016wvh},\cite{Qaemmaqami:2017jxz},\cite{Ling:2017jik}, actually it needs more investigations and also calculations in this model and other higher derivative gravity models to access a deeper understanding of this natural phenomena.
\section*{Acknowledgment}
I would like to thank Ali Naseh for proposing this topic and collaboration in initial stage of the work, Seyed Farid Taghavi for useful hints and discussions and Siavash Neshatpour for comments on manuscript. I am grateful to Ahmad Shirzad And Mohsen Alishahiha for encouragement and support. I also acknowledge the use of M. Headrick’s excellent Mathematica package “diffgeo”. I would like to thank him for his generosity.


\end{document}